\documentstyle[12pt,aaspp4]{article}
\newcommand{\be}{\begin{equation}}
\newcommand{\ee}{\end{equation}}
\newcommand{\ba}{\begin{eqnarray}}
\newcommand{\ea}{\end{eqnarray}}
\begin{document}
\title{Reconnection via the Tearing Instability}

\author{ A. Lazarian\altaffilmark{1} and Ethan T. Vishniac\altaffilmark{2}}
\altaffilmark{1}{Princeton University Observatory, Princeton, NJ 08544}

\altaffilmark{2}{Department of Astronomy, University of Texas, Austin TX
78712}

\begin{abstract}
We discuss the role of tearing instabilities in magnetic reconnection.
In three dimensions this instability leads to the formation of strong 
Alfvenic waves that remove plasma efficiently from the reconnection 
layer. As a result the instability proceeds at high rates while staying close
to the linear regime. Our calculations show that for a resistive fluid the reconnection speed scales as $V_A Re^{-3/10}$, where $V_A$ is the Alfven velocity and $Re$ is the magnetic Reynolds number. In the limit
of vanishing resistivity, tearing modes proceed at a non-zero rate, driven
by the electron inertia term, giving rise to a reconnection speed
$\sim V_A (c/\omega_pL_x)^{3/5}$, where $\omega_p$ is the plasma
frequency and $L_x$ is the transverse scale of the reconnection layer.
Formally this solves the problem of fast reconnection,  but in practice 
this reconnection speed is small.
\end{abstract}


\section{Introduction}

Reconnection of magnetic field lines is a problem that has been hotly
debated for the last forty years. Its critical importance stems from the
fact that understanding the origin and evolution of large
scale magnetic fields is impossible without a knowledge of the
mobility and reconnection of magnetic fields. Standard dynamo
theories employ the concept of turbulent diffusion to circumvent
the problems associated with the high conductivity of astrophysical
plasmas (see Parker 1979, Moffat 1978, Krause \& Radler 1980). 
Without some sort of enhanced diffusion flux freezing would be an 
excellent approximation
to the motion of the magnetic field in a highly conducting fluid.
To change magnetic field topology, to form large scale fields
from the small scale loops produced by turbulent motions, one
needs to invoke Ohmic diffusion in some way. 
As this is usually very slow in astrophysical contexts, 
the turbulent diffusion paradigm appeals to the notion that whenever field lines 
are properly intermixed, Ohmic dissipation may be enhanced by introducing
a very small magnetic correlation scale.

However, this concept has been widely criticized as
ill-founded (e.g. Parker 1992, Zweibel 1998). Strong large
scale magnetic fields should prevent
magnetic fields of opposite polarity from intermixing by turbulent
hydrodynamic motions. Both numerical and analytic studies 
(see \cite{CV91}, \cite{KA92}) confirm that the traditional (\cite{RSS88})
theory of kinematic dynamos is seriously and fundamentally flawed.
On the other hand, observations of the solar corona and chromosphere
seem to show that reconnection often takes place at speeds of
$\sim 0.1V_A$ (cf. Dere 1996, Innes et al. 1997 and references contained
therein).  Evidently at least some astrophysical plasmas can undergo 
reconnection on short time scales.

Unfortunately, current proposed solutions to this problem are not
satisfactory. The widely cited work by Parker (1992) assumes that 
the galactic dynamo depends on reconnection in the galactic halo,
where it is driven by cosmic rays.  This leaves the problem of
reconnection in other astrophysical contexts.  Moreover
it is far from obvious that reconnection in the galactic halo 
can provide the basis for an efficient galactic dynamo.

One of us (\cite{V95a}) has proposed
that rapid reconnection of magnetic flux follows from
the formation of intense flux tubes in a turbulent
plasma. However, this assumes that reconnection is initially 
rapid enough to allow the formation of such structures in a small number of
dynamical time scales and that the plasma has a high $\beta$
so that the magnetic field can be distributed intermittently.
The first assumption may reasonably be regarded as ignoring
the question of fast reconnection, since the conditions for flux
tube formation are only marginally weaker than the conditions for
fast reconnection when the field is not intermittent.  Even
granting these assumptions, while flux tubes formation 
certainly takes place in the solar convection
zone and may also be relevant to accretion discs (\cite{V95b}),
it does not seem to provide a universal solution (Lazarian
\& Vishniac 1996). For instance, ambipolar diffusion can
infiltrate material into flux tubes and suppress turbulent
pumping.  Estimates of the reconnection rate
of such flux tubes using the Sweet-Parker reconnection
process (\cite{S58}, \cite{P57}) do not guarantee reconnection
in less than an eddy turn over time under these conditions. 

In our earlier paper (Vishniac \& Lazarian 1998) we studied 
the role of ambipolar diffusion in reconnection adopting
a simple two dimensional Sweet-Parker (\cite{S58}, \cite{P57})
geometry. The reconnection speed was shown to be 
enhanced, but the level of the enhancement was neither sufficient
to account for efficient turbulent pumping nor to satisfy 
the requirements of the galactic dynamo.  In addition, this
process is not relevant to reconnection in the solar
chromosphere and corona. These considerations motivate our current
study of enhanced reconnection.

In this paper we calculate reconnection speeds
in the presence of the tearing instability, when the three dimensional
structure of the reconnection region is properly accounted for.

The tearing mode instability, which is a particular type
of resistive instability,  was quantitatively described by 
Furth, Killeen, and Rosenbluth (1963).  It is a generic instability
which has been cited in many contexts, and in particular, has
often been discussed
as a means of explaining reconnection rates associated with 
solar flaring (e.g.
Bulanov, Sakai, \& Syrovatskii 1979, Dere 1996, Glukhov 1996).

One of the problems with the customary treatment of the tearing modes is that
in a two dimensional treatment, after a short period of linear growth, they enter
a nonlinear stage that depends on the still unclear, but probably
slow, evolution of magnetic islands formed during the linear stage 
(see \cite{ML84}). In this paper we note that the case
of island formation is really singular and that in the generic case,
where the opposing field lines are not perfectly anti-parallel,
magnetic islands do not form.
Instead, strong Alfvenic waves are produced, which
carry fluid away from the reconnection zone  so that the
instability starts again for new portions of magnetic 
flux. 

The speed of reconnection is determined by the most rapidly growing tearing
modes, which reach the end of their linear growth in the time required to
eject magnetic flux in the transverse direction.  Slower modes do not have 
time to develop as the fluid and the flux are carried away.  Consequently,
the instability just barely reaches the nonlinear stage, and a linear analysis
of the problem is adequate for a qualitative analysis.

A peculiar feature of tearing modes is that they persist as fluid
resistivity vanishes. This formally solves the problem of ``fast''
dynamo, i.e. the dynamo in fluid with resistivity tending to zero,
but leaves us with the problem of real world astrophysical dynamo
as the reconnection rates that we find are small.

In what follows we discuss the physics of the tearing instability (\S
2), reconnection in resistive fluids (\S 3), and very highly conducting 
fluids (\S 4).  A summary of our results is presented in section 5.

\section{Tearing modes}

The tearing instability arises from the decoupling of magnetic field lines
from the fluid. This can be due to non-zero resistivity, electron inertia or
electron shear viscosity. If two opposite magnetic flux regions are 
brought into contact,
the instability forms magnetic `islands', as shown in Fig.~1. 
In the presence of a shared component of magnetic field perpendicular to
the plane of the figure it is easy to see that 
nonlinear Alfven waves rather than islands are formed. 
Although the projection
of these waves to the plane of Fig.~1 looks like islands, the dynamics
of the waves is radically different from that of islands. The latter
stagnate as the instability enters its nonlinear regime, while 
the former efficiently drive fluid away from the reconnection zone as the
Alfven waves propagate away from the reconnection zone.  We note that
the formation of islands also suggests the illusion that isolated loops
of magnetic flux can leave the reconnection zone in any direction, ejected
by a local pressure excess, whereas the reality is that any such motion
would involve radical distortions of the magnetic fields, and can
be ruled out on energetic grounds.

The classic study by \cite{FKR63} (hereinafter FKR) 
concluded that the tearing mode growth rate at
low wave numbers is
\begin{equation}
\gamma= \left({S\over\alpha}\right)^{2/5} {\eta\over a^2},
\label{gamma}
\end{equation}
where 
\begin{equation}
S\equiv {V_A a\over\eta},
\end{equation}
\begin{equation}
\alpha\equiv ka,
\end{equation}
$\eta$ is the resistivity, $a$ is the current sheet thickness, and
$k$ is the transverse wavenumber of the tearing mode.  This result has
been confirmed by all subsequent work.  We see that the fastest
growing modes are those with the longest transverse wavelength.

There is a controversy, however, on the the minimum wavenumber 
of the growing modes. FKR conclude that 
the instability only exists for
\begin{equation}
S^{-1/4}<\alpha<1.
\end{equation}
Using the fastest growing mode, which also gives the fastest reconnection
speed and so can be assumed to dominate transport, we get
\begin{equation}
\gamma \approx S^{1/2} {\eta\over a^2},
\label{gamma2}
\end{equation}
On the other hand, Van Hoven \& Cross (1971) (hereinafter VHC) point out that
this solution assumes infinite magnetic fields far from the origin. 
They solve numerically for the minimum transverse 
wavenumber and the fastest growing modes and 
suggest another scaling, equivalent to:
\begin{equation}
\alpha>S^{-3/7},
\end{equation}
so that
\begin{equation}
\gamma\approx S^{4/7} {\eta\over a^2}. 
\label{gammaalt}
\end{equation}
The difference between VHC and FKR can be understood in physical
terms.  Sharp gradients in the former treatment allow the instability to proceed up
to the largest scales.  Ultimately, the limiting scale is set by the
condition that $kV_A>\gamma$. Combining this criterion with equation
(\ref{gamma}) we recover equation (\ref{gammaalt}) for the maximum
growth rate.  (VHC actually quote an exponent of $0.57$.  Here we have
taken the liberty of replacing that with the numerically indistinguishable,
but physically motivated value of $4/7$.) 
On the other hand, in FKR's treatment gradients are artificially
reduced so that the instability stops at smaller scales.  
This suggests that the
work of VHC is more realistic, as gradients sharpen in the
course of reconnection with $V_{rec}<V_A$.  In what follows we will adopt Eq.
(\ref{gammaalt}), and show the results of using Eq. (\ref{gamma2}) in the 
Appendix.
We note that Bulanov et al (1979) proposed simply using $L_x$, the transverse
scale of the reconnection region, as the limiting transverse wavelength.
For our purposes this is exactly the same as using the results of VHC.

If the fluid conductivity is high, electron inertia and electron shear 
viscosity can generate the tearing instability (see Manheimer \& 
Lashmore-Davies 1984).
A simple replacement $\eta\rightarrow (m_ec^2\gamma)/(n_e e^2)$ should be
used to account for electron inertia, while a more elaborate treatment
is required for electron shear viscosity.

\section{Tearing Reconnection}

Imagine the usual situation for Sweet-Parker reconnection.
We have two volumes containing magnetic fields with strongly
differing directions.  To linear order and disregarding viscosity 
effects\footnote{The shared component of magnetic field decreases
the viscosity of the plasma in the reconnection layer (see Glukhov 1996).} 
we can ignore the magnetic field component that is shared by both regions,
so that the problem reduces to the one considered by FKR,
except that the shared field component causes the unstable intermediate
layer to shed matter out both ends at a speed $\sim V_A$ and 
with a local shear $\sim V_A/L_x$. Magnetic tension 
pulls reconnected magnetic field lines with the entrained conducting
fluid out of the reconnection zone.

This instability will be suppressed if the transverse shear exceeds
$\gamma$.  When the instability exists it mixes together magnetic fields
with opposing polarities, thereby increasing the current layer thickness
$a$ (see Fig.~1).  Therefore we have a dynamic equilibrium when (\cite{BSS79})
\begin{equation}
{V_A\over L_x}\approx \gamma,
\label{eq:tear}
\end{equation}
where $\gamma$ is the maximum growth rate, attained by the longest transverse
wavelength modes.  The resulting reconnection speed is $\sim a\gamma$, 
where we take $a$ as a characteristic
scale of the amplitude of the most rapidly growing tearing perturbations
(\cite{BSS79}). 
Expression (\ref{gammaalt}) implies that 
\begin{equation}
{a\over L_x}= S^{-3/7}.
\end{equation}
If we define the magnetic Reynolds number as
\begin{equation}
Re\equiv {L_xV_A\over\eta}={L_x\over a} S,
\end{equation}
then we obtain
\begin{equation}
S=Re^{7/10}.
\end{equation}
This gives
\begin{equation}
a=L_x Re^{-3/10},
\label{a}
\end{equation}
and the reconnection speed is
\begin{equation}
V_{rec}=a\gamma=V_A Re^{-3/10}~~~.
\label{Vrec}
\end{equation}
This rate is significantly faster than conventional Sweet-Parker reconnection speeds
although still small when $Re$ is large, as it is in most astrophysical
plasmas.

We note, that although $\gamma$ scales as $\eta^{3/5}$ the scaling
of reconnection rate is different. This is the consequence of
the fact that the thickness of current layer is also a function of
$\eta$.
Our treatment is only self-consistent if $kL_x\ge 1$.  Since 
\begin{equation}
ka=S^{-3/7},
\end{equation}
this is the same as requiring
\begin{equation}
L_x\ge a S^{3/7}=L_x Re^{-3/10} Re^{3/10}=L_x
\end{equation}
which is always true.  We see from this that adopting an upper
limit of $L_x$ for the transverse wavelength is equivalent to using
VHC's result.

\section{Reconnection in High Conductivity Plasmas}

The tearing instability persists in the limit $\eta\rightarrow 0$. In
this case the electron inertia term substitutes for the resistivity.
This was shown using the Vlasov equation (see  Hoh 1966) and confirmed 
by Cross and Van Hoven (1976)  using magnetohydrodynamic theory.

There is some controversy over whether or not the electron inertia term
can actually change the topology of magnetic field lines 
(see Shivamogy 1997). However, this argument 
seems to be of purely academic interest. The development of the
tearing instability in the presence of the electron inertia term
results in sharp current gradients and therefore any residual
fluid resistivity is sufficient to enable the actual reconnection
of the field lines.

In other words, if the conductivity is high the term $\omega(c/\omega_p)^2$,
where $\omega_p$ is the plasma frequency and $\omega\approx \gamma$,
should be substituted instead of $\eta$ in
the expression for the growth rate (\ref{gamma2}):
\begin{equation}
\gamma \approx {V_A \over a}\left({c\over\omega_p a}\right)^2~~~.
\end{equation}

Since $\gamma\approx V_A/L_x$ this implies
\begin{equation}
a\approx L_x^{2/5}\left({c\over\omega_p}\right)^{3/5}~~~.
\end{equation}
As a result the reconnection rate $V_{rec}=a\gamma$ is
\begin{equation}
V_{rec}\approx V_A \left({c\over L_x\omega_p}\right)^{3/5}~~~,
\end{equation}
which constitutes the minimal reconnection rate achievable in plasma
with $\eta\rightarrow 0$.

For typical parameters of the cold interstellar medium
\be
\omega_p\approx 10^{3.5} (n_e/0.003~{\rm cm}^{-3})^{1/2}~{\rm s}^{-1}~~~,
\ee
where $n_e$ is electron density. Therefore 
\be
{c\over L_x \omega_p}\approx 3\times
10^{-12} (n_e/0.003~{\rm cm}^{-3})^{-1/2}
(L_x/1~{\rm pc})^{-1}~~~,
\ee
and reconnection velocities will be $\sim 10^{-7}$ of the
Alfven speed . We see that although $V_{rec}$ does
not go to zero with the conductivity, the minimum reconnection 
speed is slow indeed. 

Finally, we note that if tearing modes dominate reconnection, 
it may be seen that the ratio of the reconnection rates in the 
collisionless and resistive regimes is 
\begin{equation}
\frac{V_{rec, inertia}}{V_{rec, coll}}\approx 
\left[{V_A t_{col}\over L_x}\right]^{3/10}~~~,
\end{equation}
where $t_{col}$ is the electron collision time. Naturally,
the higher the magnetic field, the more important the electron 
inertia term becomes.  However, the most important point is
that the electron inertia term is important only when electrons
do not collide in a shearing time $L_x/V_A$.  In practice this
is equivalent to saying that this term is almost never
important in an astrophysical context.

\section{Summary and Conclusions}

In this paper we have shown that three dimensional tearing modes
in resistive fluids lead to 
reconnection speeds that scale as the magnetic Reynolds number
to the minus three tenths power, compared to the scaling
$Re^{-1/2}$ in the Sweet-Parker model. In addition, we have
found that unlike standard
Sweet-Parker reconnection, reconnection involving tearing modes persists
as the fluid resistivity tends to zero while electron inertia
drives tearing modes. This formally solves the problem of ``fast''
dynamo, but is of marginal assistance to actual astrophysical dynamos
as the tearing reconnection is not fast.

The treatment so far ignores viscosity effects. Electron viscosity
can initiate tearing modes on its own. The corresponding term in
Ohm's law is proportional to the second derivative of the current.
In the case of tearing modes driven by electron inertia this term
may become important when sharp current gradients form.
The energy deposited is dissipated by viscous dissipation.

In a recent study Glukhov (1996) has shown that viscosity
plays an important role when the current sheet is neutral, i.e.
there is no magnetic field in the direction of current. We believe
that such situations are rather singular and, in general, there
will be a component of magnetic field along current.
Moreover, a neutral current sheet is likely to be subjected 
to a strong interchange instability and therefore its response 
to slower tearing instability is only of academic interest.

In the present paper we do not treat the highly controversial case of ``forced
reconnection'' involving tearing modes (see Hassam 1992) and follow 
the line of reasoning adopted in FKR and later studies
(e.g. Kulsrud \& Hahm 1982). 

Finally, we note that our conclusions differ quite dramatically from
those of Strauss (1988), who claimed that tearing modes would lead
to fast reconnection, i.e. $V_{rec}\sim V_A$ on the basis of a
weakly nonlinear analysis of tearing mode interactions.  However,
there is less to this disagreement than would appear at first
glance.  We agree that tearing modes will grow to the point
of marginal nonlinearity.  We do not agree that this 
implies fast reconnection.  Instead, global
constraints play a dominant role in setting $V_{rec}$, as they
do in the usual Sweet-Parker argument without tearing modes.

We conclude that by themselves tearing modes do not look like a panacea for
the problems of the astrophysical dynamo. Indeed, although
the enhancement of reconnection speeds is substantial in 
numerical terms, the consequent reconnection rates are not sufficient
to support contemporary dynamo theories. Reconnection
in the presence of the tearing modes is still too slow. 

Is this a crisis? Probably not.  Our present model is still too 
simple to treat realistic reconnection geometries. In our next paper
we will show that the reconnection speeds are substantially enhanced in the
presence of MHD turbulence. 

\acknowledgements

We are grateful to E. Zweibel, B. Draine and R. Kulsrud for a 
series of helpful discussions.  This work was supported in
part by  NASA grants NAG5-2858 (AL), NAG5-2773 (ETV), and
NSF grant AST-9318185 (ETV).  ETV
is grateful for the hospitality of MIT and
the CfA during the completion of this work.

\appendix

\section{FKR Treatment}

FKR assume that the magnetic field strength increases linearly
with distance to the neutral line, while a more realistic 
structure of the background magnetic field is discussed by VHC.
In the text we use the latter result.
What is more important to our analysis is that
tearing modes happen with very similar growth rates in quite different 
magnetic configurations. Therefore our results are robust.
Below we present the reconnection speeds
based on the work of FKR.

In this case instead of Eq.~(\ref{eq:tear}) we get
\be
\frac{V_A}{L_x}\approx S^{1/2}\frac{\eta}{a^2}~~~,
\ee
which means that
\be
\frac{a}{L_x}\approx S^{-1/2}~~~.
\ee
Therefore $Re=(L_x/a)S\approx S^{3/2}$, which means that
\be
a\approx L_x Re^{-1/3}
\ee
and
\be
V_{rec}\approx V_A Re^{-1/3}~~~,
\ee
which should be compared with Eq.~(\ref{Vrec}).

In the limit of negligible resistivity
\be
\gamma \approx {V_A L_x\over a^2}\left({c\over\omega_p a}\right)^2~~~,
\ee
and by equating $\gamma$ to $V_A/L_x$ we get
\be
a\approx L_x^{3/5}\left({c\over\omega_p}\right)^{2/5}~~~.
\ee

Finally,
\be
V_{rec}\approx V_A \left({c\over L_x\omega_p}\right)^{2/3}~~~,
\ee
which differs insubstantially from our estimate (\ref{Vrec}).

\clearpage
\begin{figure}
\begin{picture}(441,216)
\includegraphics{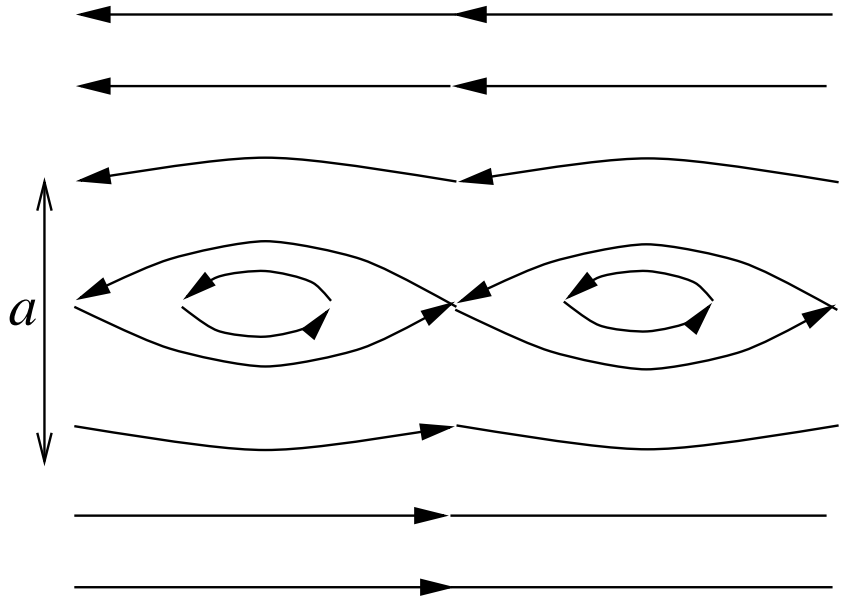}
\end{picture}
\figcaption{A schematic of a reconnection region.}
\end{figure}


\begin{thebibliography}{}
%\bibitem[Brandenburg \& Zweibel\ 1995]{BZ95}
%Brandenburg, A., \& Zweibel, E.G.\ 1995, \apj, 448, 734
\bibitem[Bulanov, Sakai, \& Syrovatskii 1979]{BSS79} Bulanov, S.V., Sakai, J., \& Syrovatskii, S.I. 1979,
Sov. J. Plasma Phys. 5(2), 157
\bibitem[Cattaneo \& Vainshtein\ 1991]{CV91}
Cattaneo, F., \& Vainshtein, S.I.\ 1991, \apj, 376, L21
%\bibitem[]{} Craig, I.J.D., \& Mc Clymont, A.N. 1991, ApJ, 371, L41
\bibitem[]{} Cross, M.A., \& Van Hoven, G. 1976, Phys. Fluids,
19, 1591
%\bibitem[Dalgarno\ 1958]{D58} Dalgarno, A.\ 1958, Phil. Trans.
%Roy. Soc., Ser. A, 250, 426
\bibitem[]{} Dere, K.P. 1996, ApJ, 472, 864
\bibitem[Furth, Killeen, \& Rosenbluth\ 1963]{FKR63} Furth, H.P.,
Killeen, J., \& Rosenbluth, M.N.\ 1963, Phys. Fluids, 6, 459 (FKR)
\bibitem[]{} Glukhov, V. 1996, \apj, 469, 936
%\bibitem[Gruzinov \& Diamond\ 1994]{GD94}
%Gruzinov, A., \& Diamond, P.H.\ 1994, Phys. Rev. Lett., 72, 1651
\bibitem[]{} Hassam, A.B. 1992, \apj, 399, 159
\bibitem[]{} Hoh, F.C. 1966, Phys. Fluids, 9, 277
%\bibitem[]{} Kazantsev, A.P. 1968, Soviet Phys. - JETP, 26, 1031
\bibitem[Innes et al.\ 1997]{IIAW97}
Innes, D.E., Inhester, B., Axford, W.I., \& Wilhelm, K.\ 1997, \nat, 386, 811
\bibitem[]{} Krause, F., \& Radler, K.H. 1980, Mean-Field 
Magnetohydrodynamics
and Dynamo Theory, Oxford: Pergamon Press
\bibitem[Kulsrud \& Anderson\ 1992]{KA92} 
Kulsrud, R.M., \& Anderson, S.W. \ 1992, \apj, 396, 606
\bibitem[]{} Kulsrud, R.M., \& Hahm, T.S. 1982, Physica Scripta,
T2/2, 525
%\bibitem[]{} Lazarian, A. 1992, A\&A, 264, 326
\bibitem[Lazarian \& Vishniac\ 1996]{LV96} 
Lazarian A., \& Vishniac, E.T.\ 1996, in Polarimetry of the 
Interstellar Medium, eds. W.G.~Roberge and D.C.B.~Whittet, ASP 97, 537
%\bibitem[Longair\ 1994]{L94} Longair, M. 1994, High Energy Astrophysics, 
%(New York: Cambridge University Press) 163
\bibitem[ Manheimer \&  Lashmore-Davis\ 1984]{ML84} Manheimer, W.M., 
\& Lashmore-Davis, C. 1984,
MHD Instabilities in Simple Plasma Configuration, Naval
Research Laboratory, Washington
\bibitem[]{} Moffatt, H.K. 1978, Magnetic Field Generation in Electrically
Conducting Fluids Cambridge: Cambridge University Press
\bibitem[Parker\ 1957]{P57}Parker, E.N.\ 1957, J. Geophys. Res., 62, 509
\bibitem[]{} Parker, E.N. 1979, {\it Cosmical Magnetic Fields}, Oxford: 
Clarendon Press
\bibitem[Parker\ 1992]{P92}Parker, E.N.\ 1992, \apj, 401, 137
%\bibitem[]{} Petschek, H.E. 1964, {\it The Physics of Solar Flares}, AAS-NASA
%Symposium, NASA SP-50 (ed. W.H. Hess), Greenbelt, Maryland, p.~425
\bibitem[Ruzmaikin et al.\ 1988]{RSS88}
Ruzmaikin A.A., Shukurov A.M. \& Sokoloff D.D. 1988 Magnetic fields
of Galaxies, Kluwer, Dordrecht
\bibitem[]{} Shivamogy, B.K. 1997, Journal of Plasma Physics, 58, 329
%\bibitem[]{} Spitzer L., Jr. 1978, Physical Processes in the Interstellar
%Medium, Wiley-Interscience Publ., New York
\bibitem[Strauss \ 1988]{S88} Strauss, H.R.\ 1988, \apj, 326, 412
\bibitem[Sweet\ 1958]{S58} Sweet, P.A.\ 1958, in IAU Symp. 6, Electromagnetic 
Phenomena in 
Cosmical Plasma, ed. B. Lehnert (New York: Cambridge Univ. Press), 123
\bibitem[Van Hoven \& Cross\ 1971]{VHC71} Van Hoven, G., \& Cross, M.A.\ 1971,
Phys. Fluids, 14, 1141 (VHC)
\bibitem[Vishniac\ 1995a]{V95a} Vishniac, E.T.\ 1995a, \apj, 446, 724
\bibitem[Vishniac\ 1995b]{V95b} Vishniac, E.T. 1995b, {\it ApJ}, 451, 816
\bibitem[]{} Vishniac, E.T., \& Lazarian, A. 1998, \apj, (submitted)
\bibitem[]{} Zweibel, E. 1998, Physics of Plasmas, v. 5, 247 
\end{thebibliography}
\end{document}